\documentstyle[aps,prb,preprint]{revtex}

\begin{document}
\draft
\title{Numerical renormalization group study of the $1D$ $t-J$ model}
\author{Liang Chen$^1$ and S. Moukouri}
\address{D\'epartement de physique and Centre de recherche en physique du
solide,\\
Universit\'e de Sherbrooke, Sherbrooke, Qu\'ebec, Canada J1K 2R1}
\date{\today}
\maketitle

\begin{abstract}
The one-dimensional ($1D$) $t-J$ model is investigated using the density
matrix renormalization group (DMRG) method. We report
for the first time a generalization
of the DMRG method to the case of arbitrary band filling and prove a theorem
with respect to the reduced density matrix that accelerates the
numerical computation. Lastly, using the extended DMRG method, we present
the ground state electron momentum distribution, spin and
charge correlation functions. The $3k_F$ anomaly of the momentum
distribution function first discussed by Ogata and Shiba
is shown to disappear as $J$ increases.
We also argue that there exists a density-independent $J_c$ beyond which the
system becomes an electron solid.
\end{abstract}

\pacs{71.10.+x,71.27.+a,75.10.Jm}

\section{INTRODUCTION}

For a number of years, the more so since
the discovery of high-$T_c$ superconductors\cite{Bednorz}, the study
of strongly correlated electron systems has been a major theme of theoretical
condensed matter physics.
The $t-J$ model is one of the simplest of such models\cite{Anderson}.
Although high-$T_c$ cuprates
are at least two-dimensional ($2D$) systems, it is also relevant to
fully understand
the one-dimensional ($1D$) model. The Hamiltonian for the $t-J$
model in one dimension can be written in the subspace of no doubly occupied
sites as

\begin{equation}
H=-t\sum_{i\sigma }(c_{i\sigma }^{\dagger }c_{i+1\sigma }+c_{i+1\sigma
}^{\dagger }c_{i\sigma })+J\sum_i({\bf S}_i{\bf \cdot S}_{i+1}-\frac 14%
n_in_{i+1}),  \label{hamiltonian}
\end{equation}
where $c^\dagger_{i\sigma}$ and $c_{i\sigma}$ are the creation and
annihilation operators for an electron at lattice site $i$ with spin $\sigma$,
while ${\bf S}_i$
and $n_i$ are the corresponding spin and electron number operators.
This model has been solved exactly only for $J\rightarrow 0$, where it is
equivalent to the $U\rightarrow \infty $ Hubbard model, and at the
supersymmetric point\cite{Bares} $J=2t$. In both cases the ground state at
arbitrary density belongs to a broad class of interacting Fermi systems
known as Luttinger liquids\cite{Haldane}, which exhibit power-law decay of
correlation functions with exponents characterized by a single parameter\cite
{Kawakami}. Additionally, for very large $J/t$, the attractive Heisenberg
interaction term dominates the kinetic energy and the model phase separates%
\cite{Phase}$^{,}$\cite{VMonte}. The level of understanding of this model is
derived mostly from small cluster exact diagonalizations\cite{Phase},
variational Monte Carlo methods\cite{VMonte} and finite temperature Monte
Carlo simulations for relatively larger system sizes\cite{Assaad}.
There still lurks, however,
the question of whether the thermodynamic limit has been
reached or
not. In this paper we use the density matrix renormalization group
(DMRG) method\cite{White} to study the ground-state properties of the
$t-J$ model in the thermodynamic limit.

The text is organized as follows. In Sec. II, we review the formulation of
DMRG and its extension to arbitrary band-filling
. In Sec. III, we discuss our numerical results. We
present ground-state static spin and charge correlation functions as
well as electron momentum distribution functions.
Finally in Sec. IV we summarize our results.

\section{DMRG FORMULATIONS}

The DMRG technique was developed by White\cite{White} in 1992. It
leads to highly accurate results for systems much larger than
those which can be solved by exact diagonalization. The DMRG
allows a systematic reduction of the Hilbert space to
the basis states most
relevant to describe a given eigenstate ($e.g.$ the ground state) of a large
system. This is to be contrasted with previous real
space renormalization techniques in which the lowest states are kept.
A general iteration step of the method for open
boundary conditions proceeds as follows:
$a)$ The effective Hamiltonian defined for
the superblock $1+2+2'+1'$ (where the block 1 and $1'$ come from previous
iterations and 2 and $2'$ are new added ones) is diagonalized to obtain the
ground-state wave function
$|\psi >$ (other states could be also kept). $b)$ The reduced
density matrix of blocks $1+2$: $\rho _{i,i^{\prime }}=\sum_j\psi _{ij}\psi
_{i^{\prime }j}^{*}$ is constructed, where $\psi _{ij}=<i\otimes j|\psi >$,
the states $|i>$ ($|j> $) belongs to the Hilbert space of blocks 1 and 2 ($1'$
and $2'$). The eigenstates of $\rho $ with the highest eigenvalues
(equivalent to the most probable states of blocks $1+2$ in the ground state of
the superblock) are kept up to a certain cutoff. $c)$ These states form a
new reduced basis in which all the operators have to be expanded and the
block $1+2$ is renamed as block 1. $d)$ A new block 2 is added (one site in
our case) and the new superblock $(1+2+2'+1')$ is formed as the direct product
of the states of all the blocks (the blocks $1'$ and $2'$ are identical to
blocks 1 and 2 respectively). The method has been successfully applied to
problems such as the Haldane gap of spin-1 chains, critical exponents of
spin-$\frac 12$ chains, the $1D$ Kondo-insulator and two-chain Hubbard model%
\cite{DmrgA}.

In our study of the $1D$ $t-J$ model we have used the infinite-size version
of the above iteration scheme to reach the thermodynamic limit. One
immediately realizes, however,
 there is a problem in keeping the electron density
fixed in the iteration process since we insert only two sites at each
iteration (which only makes the half-filling and quarter-filling cases
invariant). To get around this, we construct the reduced density matrix
from two ground states that bracket the desired density.
To be more specific, if our
desired electron density is $n$ and the current superblock lattice size is $%
N $, then we can always find two nearest integers $N_1$ and $N_2$ such that $%
N_1\leq nN\leq N_2$. Assuming $|\psi (N_1)>$ is the ground state wavefunction
of $N_1$ electrons and $|\psi (N_2)>$ is that of $N_2$ electrons, we build
the reduced density matrix by the following weighting procedures:
\begin{equation}
\left\{
\begin{array}{l}
\rho _{i,i^{\prime }}=\rho _1\sum_j\psi _{ij}(N_1)\psi _{i^{\prime
}j}^{*}(N_1)+\rho _2\sum_j\psi _{ij}(N_2)\psi _{i^{\prime }j}^{*}(N_2) \\
nN=\rho _1N_1+\rho _2N_2 \\
1=\rho _1+\rho _2
\end{array}
\right.
\end{equation}
It is clear that the above construction ensures the desired constant
band filling at every iteration. The success of this construction inherently
requires that the ground state be homogeneous. Therefore one
should expect failure when the system goes into the phase separation regime.
However, phase separation can still be studied using the finite size
version of DMRG\cite{White} which does not require translational
invariance. It is perhaps worthwhile to remark that in our computer
program
we have generally targeted three states since the ground state has
two-fold degeneracy when the electron number is odd.

The correlation functions are calculated when the iteration has converged.
Typically we start to measure them after the superblock size reaches 40
lattice sites. More specifically we measure the correlation between operators
in the middle of the superblock. For example, the nearest-neighbor correlation
is measured between sites 20 and 21 in the superblock size 40; while the
second-nearest-neighbor correlation is measured between sites 20 and 22 in
the next iteration (i.e when superblock size becomes 42). Similarly the
third-nearest-neighbor correlation is measured between sites 20 and 23 in
the superblock size 44. Thus as we iterate our procedure to increase the
system size we obtain correlations over increasing distances.
Finally, the desired ground-state correlations are obtained in a
similar way as the reduced density matrix, $e.g.$
the spin-spin correlation is measured in the following way
\begin{equation}
S{\bf (r}-{\bf r}^{\prime })=\rho _1<\psi (N_1)|{\bf S(r)\cdot S(r}^{\prime
})|\psi (N_1)>+\rho _2<\psi (N_2)|{\bf S(r)\cdot S(r}^{\prime })|\psi (N_2)>.
\end{equation}

Before we discuss the numerical results, let us prove
the following theorem: {\it
the eigenstates of the reduced density matrix retain their good quantum
numbers when the corresponding operator of the superblock is a direct
sum of the operators of its subsystems}.
The electron number operator $\widehat{N}$
and the total $z$-component of the spin operator $\widehat{S_z}$ are two
such operators. To prove such a theorem all we need to show is that the
reduced density matrix is block diagonal in its good quantum operator
subspace. Let us consider one subspace at a time, $e.g.$ the electron number
subspace. Then the most general form of the eigenstate of the superblock
with $L$ electrons can be written as
\begin{equation}
|\psi (L)>=\sum\limits_{l_1+l_2=L}\sum\limits_{\alpha ,\beta }c(l_1,\alpha
;l_2,\beta )|l_1,\alpha >_1\otimes |l_2,\beta >_2,
\end{equation}
where $|l_1,\alpha >_1$ stands for the basis wavefunction of block 1 with $%
l_1$ electrons, and $\alpha $ labels other quantum numbers, while $%
|l_2,\beta >_2$ represents the basis wavefunction of block 2 with $l_2$
electrons, and $\beta $ marks other quantum numbers. Thus the reduced
density matrix elements have the form
\begin{equation}
\rho _{(l_1,\alpha );(l_1^{\prime },\alpha ^{\prime
})}=\sum\limits_{l_2,\beta }c(l_1,\alpha ;l_2,\beta )\cdot c^{*}(l_1^{\prime
},\alpha ^{\prime };l_2,\beta ).
\end{equation}
However according to the construction of the wavefunction we have both $%
L=l_1+l_2$ and $L=l_1^{\prime }+l_2$ leading automatically to the
conclusion $l_1\equiv l_1^{\prime }$ . Therefore the reduced density matrix
is block diagonal in the electron number subspace. In the same way one can
show that the reduced density matrix is block diagonal in the $S_z$
subspace. We emphasize that this theorem can be explicitly implemented to
accelerate the diagonalization of the reduced density matrix
in the subspace of its good quantum numbers.

\section{NUMERICAL RESULTS}

For most of the numerical results reported here
we have kept 110 states\cite{Remark} in
blocks 1 and $1'$, the truncation error defined as $1-p(m)$ (where $p(m)$ is
the summation of the highest $m$ eigenvalues of the reduced density matrix)
is of the order of $10^{-5}$.
In Table I we list some of the ground-state energies
per site $E_g$ as a function of band filling $n$ and spin exchange coupling $%
J$. In the same table we also list the results obtained from the exact Bethe
ansatz solution at $J=2t$. As can be seen immediately from the table,
our DMRG results are highly accurate. To make it convenient for the reader
we quote the coupled Bethe ansatz equations\cite{Bares} determining
the ground-state energy at density $n$ and $J=2t$:
\begin{equation}
\int\limits_{-Q}^Q\rho (\nu )d\nu =1-n
\end{equation}
\begin{equation}
\rho (\nu )=2R(2\nu )+\int\limits_{-Q}^Q2R(2[\nu -\nu ^{\prime }])\rho (\nu
^{\prime })d\nu ^{\prime },
\end{equation}
where $R(x)$ denotes Shiba's function
\begin{equation}
R(x)=\frac 1{4\pi }\int\limits_{-\infty }^\infty d\omega \frac{e^{i\omega
x/2}}{1+e^{|\omega |}}.
\end{equation}
Then the ground-state energy is given by
\begin{equation}
E_g=2t[1-n-\pi \rho (0)].
\end{equation}

Since we are going to present our results in momentum space it is now
opportune
to discuss the way we analyse our data. First, because open boundary
conditions are used in our DMRG procedure, we obviously loose translational
invariance. To overcome this boundary effect we have done an average on the
real space correlation functions before we do the Fourier transform into
momentum space. More specifically we obtain several real-space correlation
functions of the same system by starting the measurement at different
superblock sizes. For example we start our real-space
density-density correlation $\rho\rho(r)=<n_in_{i+r}>$
measurement after the superblocks reach the sizes of 40, 42, 44, 46
and 48 sites and then take the average among them. The final Fourier
transform is done in the usual way:
\begin{equation}
\rho\rho(k)={1\over N}\sum_{l=1}^N\sum_{h=1}^N
e^{ik(l-h)}[<n_ln_h>-<n_l><n_h>],
\end{equation}
where $N$ is the largest separation available in our measurement. For the
results of this work we stop at a superblock size of 150 sites.

In Fig. 1(a) we plot the electron momentum distribution function for
quarter filled band at $J=0.1t.$ We see that this result is almost identical
to the $1D$ infinite-$U$ Hubbard model studied by Ogata and Shiba\cite{Shiba}.
This is not surprising considering the fact that the
small $J$ limit of the $t-J$
model is equivalent to the strong coupling limit of the Hubbard model. Here we
have both $k_F$ $(=n\pi/2)$ and $3k_F$
anormalies as indicated by arrows in the figure.
Unlike normal Fermi liquids, however, the $3k_F$ anomaly is a new feature
which is related to the effect of the coupled holon and spinon
representation of the normal electron\cite{Shiba}. Fig. 1(b) shows the
spin-spin correlation. There is a $2k_F$ anomaly reflecting the nature of
the antiferromagnetic exchange among nearest-neighbor electrons.
Fig. 1(c) is
the density-density correlation function. Here one observes the $4k_F$ $(=\pi)$
anomaly only.

Before discussing the effect of increasing $J$ let
us remember the two competing factors in correlation effects: $a)$
The nearest-neighbor antiferromagnetic spin exchange $J$ favors the formation
of
spin-singlet electron pairs
on nearest-neighbor sites which in turn leads to $2k_F$ spin density
wave (SDW) fluctuations. But
eventually, when $J$ is large enough, there emerges a phase separation
exhibiting antiferromagnetic SDW fluctuations.
$b)$ The kinetic energy $t$ term favors the delocalization of the electrons.
But the Pauli principle as well as on-site repulsion promote uniform
separation between electrons synonymous
with $4k_F$ charge density wave (CDW) fluctuations. On the
other hand the formation of bound singlet pairs due to $J$ term enhances
$2k_F$ CDW fluctuations.

In Fig. 2(a) it is seen that as the antiferromagnetic exchange $J$ increases,
the momentum distribution function is modified more drastically
in the region $k>k_F$ where it changes from decreasing to increasing
as a function of $k$. Notably the $3k_F$ feature has been washed out for
larger $J$. Besides, the critical exponent\cite{Kawakami}
at the Fermi wavevector $k_F$ seems to decrease as $J$ increases.
In Fig. 2(b) we have plotted the spin-spin correlation function at quarter
filling with $J/t=0.5,2,$ and 2.5. We see that at $J=0.5t$ there is a clear
$2k_F$ anomaly, however at $J=2t$ this anomaly is significantly weakened.
At $J=2.5t$ the $2k_F$ anomaly seems to have completely disappeared and
the maximum is located at $k=\pi=k_{AF}$. This is an indication of singlet
pairing between nearest-neighbor electrons.
Fig. 2(c) shows the corresponding density-density correlation.
At $J=0.5t$ there is only a $4k_F$ anomaly which indicates the system is quite
uniformly distributed. However as one increases $J$ the $4k_F$ feature
disappears and a new feature at $2k_F$ develops which reflects the
formation of nearest-neighbor pair. At $J=2.5t$ the behavior near $k=0$
has a tendency to flatten out. But the $2k_F$ feature still dominates
which shows singlet pairing.
Fig. 3 shows similar results for the band filling $n=0.8$. Again it is
seen from Fig. 3(a) that $J$ has the largest effect on the momentum
distribution
function in the regime $k>k_F$. Fig. 3(b) is very much the same as
of Fig. 2(b): as $J$ increases the maximum in the spin-spin correlation
function moves to the antiferromagnetic wavevector $k_{AF}=\pi$.
But in Fig. 3(c) the $J=3t$ density-density correlation shows a
clearer tendency towards phase separation near $k=0$. Both $2k_F$ and $4k_F$
anomalies are seen in the density correlation at intermediate
$J$ as predicted\cite{Kawakami}.

As mentioned in the previous section our DMRG procedure is not applicable
to the study of the phase separated regime. But we know from other numerical
calculations there exists a $density-dependent$ $J'_c$ indicating the onset
of a phase separation. Here we
wish to argue that there exists another
critical $J_c$ (that is larger than the $max[J'_c]$)
beyond which the system forms an $electron$ $solid$, i.e., a phase
with no hole inside but having a thin interface
at which the boundary electrons
can still evaporate into the vacuum (hole).
This $J_c$ should be independent of the band filling since
once the $electron$ $solid$ is reached there is only a single phase boundary
between the hole region and the solid, and that is
density independent. By definition,
the first instability of such an $electron$ $solid$ comes from the ability to
dissolve a single hole inside. Accordingly $J_c$ can be
estimated from finite size exact diagonalizations by comparing the
ground-state
energy of a $2N$-site Heisenberg chain with that of 2 holes in
 the $2N$-site
$t-J$ model. This has been done by Ogata et al\cite{Phase} on a 16-site
system and by Assaad et al\cite{Assaad} with path-integral Monte Carlo
simulation. Both results seem to point towards $J_c\sim 3.6t$. Further,
Yokoyama et al\cite{VMonte} have hinted the existence of such a $J_c$
in their recent variational Monte Carlo study of $t-J$ model. The same
thing is also implied in the work of Ammon el al\cite{Ammon}.

\section{SUMMARY}

We have extended the DMRG method to allow calculations at arbitrary
band filling and proven a theorem regarding the reduced density matrix
that could accelerate the numerical calculation. We then used the extended
DMRG procedure to study the $1D$ $t-J$ model. It is found that our procedure
gives highly accurate ground-state energy and correlation functions.
It is shown that the $3k_F$ anomaly first discussed by Ogata and Shiba
in the electron momentum distribution disappears as $J$ is increased.
It is argued that there exists a density-independent $J_c$ beyond which the
system forms an $electron$ $solid$.

\acknowledgements

We acknowledge the support of the Natural Sciences and Engineering Research
Council of Canada (NSERC), the Fonds pour la formation de chercheurs et
l'aide \`a la recherche from the Government of Qu\'ebec (FCAR). We are
grateful to A.-M. S. Tremblay and L.G. Caron for encouragements and reading
of the manuscript.

\begin{table}[tbp]
\caption{Ground-state energies of the $1D$ $t-J$ model calculated using DMRG
method. Exact results for the $J=2t$ case are listed as a comparison. The
last number in the brackets are the estimated value accurate at that digit.}
\begin{tabular}{|c|c|c|c|}
$<n>$ & $J/t$ & DMRG $E_g/t$ & Exact $E_g/t$ \\ \hline
0.4 & 0.5 & -0.631(8) &  \\ \hline
0.4 & 1.0 & -0.664(2) &  \\ \hline
0.4 & 2.5 & -0.803(3) &  \\ \hline
0.5 & 0.1 & -0.647(0) &  \\ \hline
0.5 & 0.5 & -0.692(0) &  \\ \hline
0.5 & 2.0 & -0.903(6) & -0.9036(4) \\ \hline
0.5 & 2.5 & -0.988(5) &  \\ \hline
0.8 & 0.1 & -0.416(2) &  \\ \hline
0.8 & 0.5 & -0.586(7) &  \\ \hline
0.8 & 2.0 & -1.246(4) & -1.2464(4) \\ \hline
0.8 & 3.0 & -1.698(7) &  \\ \hline
0.9 & 1.0 & -0.756(4) &  \\ \hline
0.9 & 2.0 & -1.322(3) & -1.3223(7) \\ \hline
0.9 & 3.0 & -1.891(9) &
\end{tabular}
\end{table}

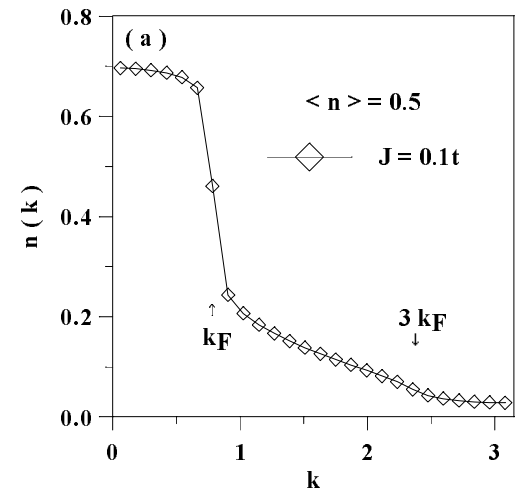
\begin{figure}
\caption{The ground state electron momentum distribution function (a),
spin-spin correlation function (b),
and density-density correlation (c) are plotted for quarter band
filling at $J=0.1t$.}
\end{figure}

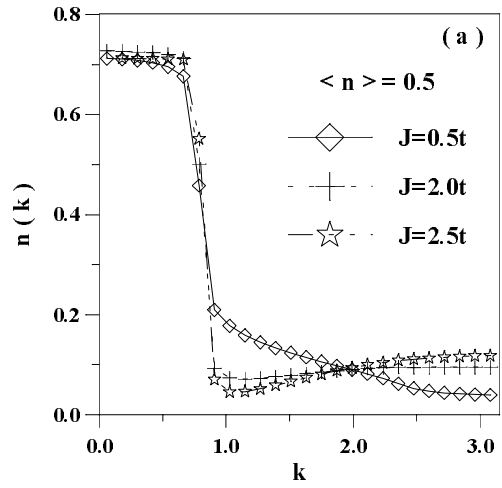
\begin{figure}
\caption{Same as that of Fig. 1 but with $J=0.5t,2t$ and $2.5t$.
}
\end{figure}

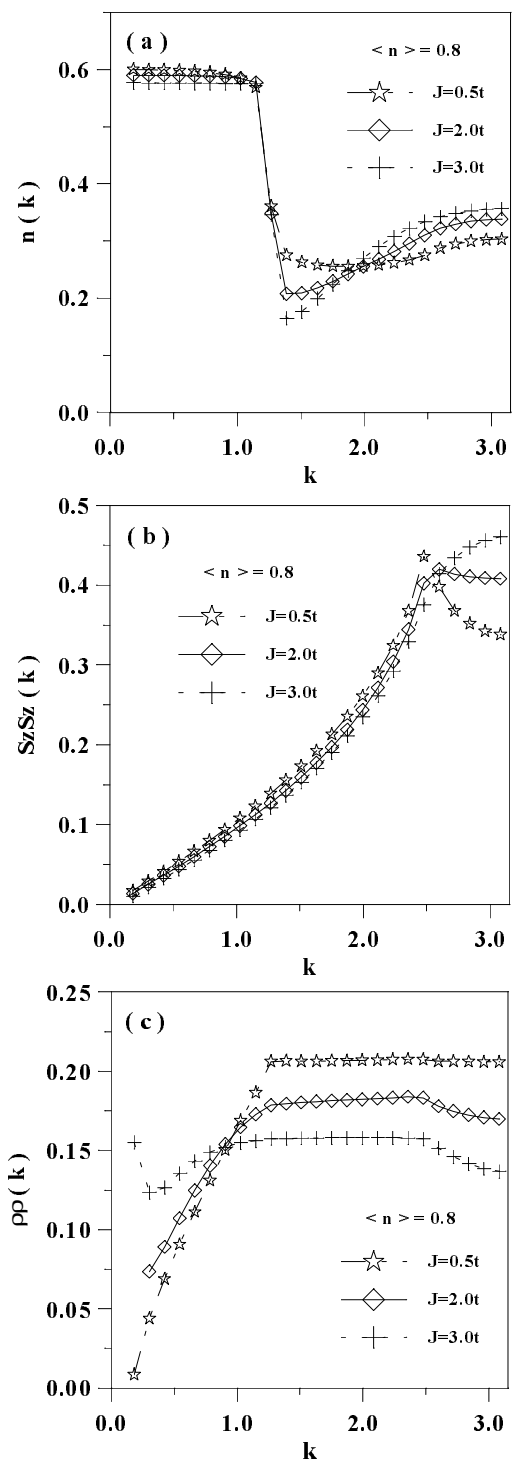
\begin{figure}
\caption{Same as that of Fig. 2 with $J=0.5t,2t$ and $3t$ but at band filling
0.8.}
\end{figure}


\begin{references}
\bibitem{Remark} The code written for this work did not implement the theorem
of block diagonalization of the reduced density matrix so we have to directly
diagonalize matrix size of $330\otimes 330$ in each iteration. However we
expect that after implementing the block diagonalization of the reduced density
matrix we can keep more states in blocks 1 and $1'$ and obtaining higher
accuracy. Work in this direction is in progress.
\end{references}
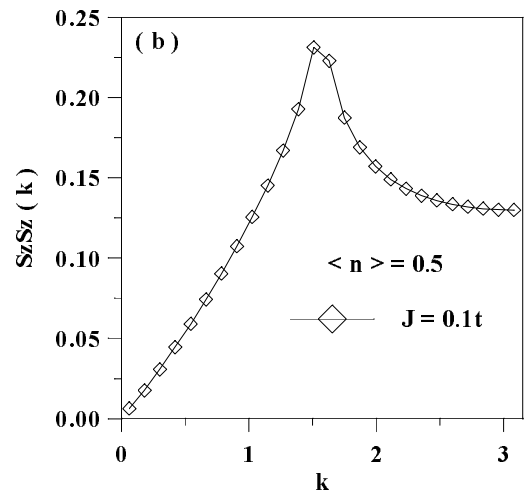
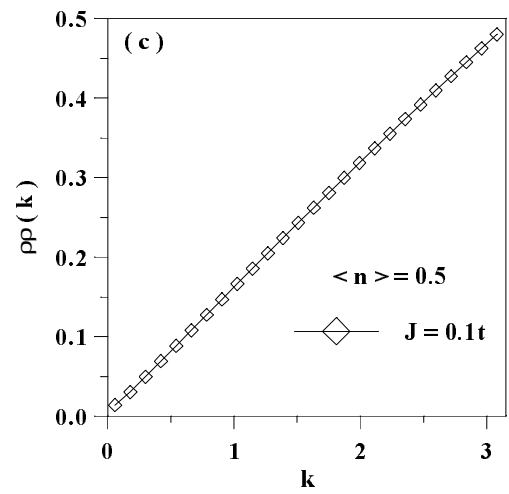
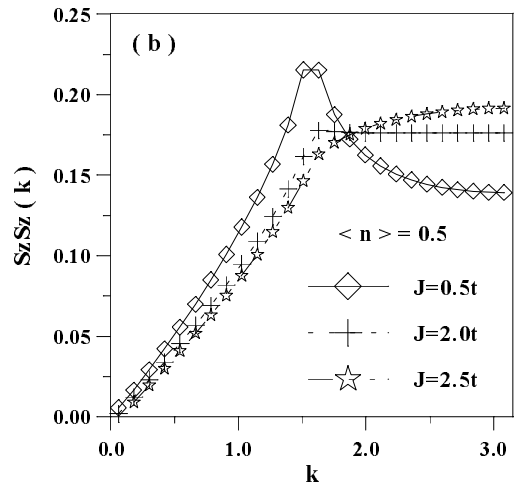
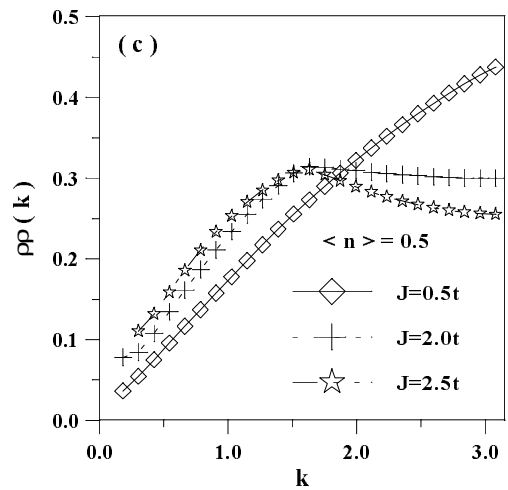
\end{document}